\documentclass{aa}
\usepackage{graphicx}
\usepackage{lscape}
\usepackage{longtable}
\usepackage{psfig}
\begin{document}
   \title{Lithium abundances of halo dwarfs based on excitation temperature. I. LTE}

   \author{A. Hosford\inst{1}
          \and
          S.G. Ryan\inst{1}
          \and
          A.E. Garc\'{i}a P\'{e}rez\inst{1}
          \and
          J.E. Norris\inst{2}
          \and
          K.A. Olive\inst{3}
          }

   \offprints{A. Hosford}

   \institute{Centre for Astrophysics Research, University of Hertfordshire,
              College Lane, Hatfield, AL10 9AB, UK\\
              \email{a.hosford@herts.ac.uk, s.g.ryan@herts.ac.uk, a.e.garcia-perez@herts.ac.uk}
         \and
             Research School of Astronomy and Astrophysics, The Australian National University, Mount Stromlo Observatory, 								Cotter Road, Weston, ACT 2611, Australia \\
             \email{jen@mso.anu.edu.au}
         \and
         		William I. Fine Theoretical Physics Institute, School of Physics and Astronomy, University of Minnesota, 			   						Minneapolis, MN 55455\\
         		\email{OLIVE@mnhep.hep.umn.edu}
             }

   \date{Received; accepted}


  \abstract
   {The discovery of the Spite plateau in the abundances of \element[][7]{Li} for metal-poor stars led to the determination of an observationally deduced primordial lithium abundance. However, after the success of the Wilkinson Microwave Anisotropy Probe (WMAP) in determining the baryon density, $\Omega_{\rm B}h^{2}$, there was a discrepancy between observationally determined and theoretically determined abundances in the case of \element[][7]{Li}. One of the most important uncertain factors in the calculation of the stellar \element[][7]{Li} abundance is the effective temperature, $\textit{T}_{\rm eff}$.}
   { We use sixteen metal-poor halo dwarfs to calculate new $\textit{T}_{\rm eff}$ values using the excitation energy method. With this temperature scale we then calculate new Li abundances for this group of stars in an attempt to resolve the \element[][7]{Li} discrepancy. }
   {Using high signal-to-noise (S/N $\approx$ 100) spectra of 16 metal-poor halo dwarfs, obtained with the UCLES spectrograph on the AAT, measurements of equivalent widths from a set of unblended {Fe} {I} lines are made. These equivalent widths are then used to calculate new $\textit{T}_{\rm eff}$ values with the use of the single line radiative transfer program WIDTH6, where we have constrained the gravity using either theoretical isochrones or the Hipparcos parallax, rather than the ionization balance. The lithium abundances of the stars are calculated with these temperatures.}
   {The physical parameters are derived for the 16 programme stars, and two standards. These include $T_{\rm eff}$, log $g$, [Fe/H], microturbulence and \element[][7]{Li} abundances. A comparison between the temperature scale of this work and those adopted by others has been undertaken. We find good consistency with the temperatures derived from the H$\alpha$ line by Asplund et al. (2006), but not with the hotter scale of Mel\'{e}ndez {\&} Ram\'{i}rez (2004). We also present results of the investigation into whether any trends between \element[][7]{Li} and metallicity or temperature are present in these metal-poor stars.}
   {}


   \maketitle
%

\section{Introduction}

\subsection{The Lithium Problem}
\label{sec:TheLithiumProblem}
The development of the Big Bang theory brought about predictions for the primordial abundances of the nuclides,   \element[][2]{H}, \element[][3]{He}, \element[][4]{He}, and \element[][7]{Li}
(\cite{Oliveetal00}, \cite{FieldsSarkar06}). These values, determined from calculations of Big Bang Nucleosynthesis (BBN) for an assumed baryon to photon ratio $\eta$, are in at least partial agreement with those primordial abundances determined from observations of \element[][2]{H}, \element[][3]{He}, \element[][4]{He}, and \element[][7]{Li} relative to \element[][1]{H}. Historically this has given constraints on $\eta$, or equivalently the baryon density fraction, $\Omega_{\rm B}h^2$, which is standard BBN's one free parameter, where $\textit{h}$ is the Hubble constant in units of 100 k$\rm m^{-1}$ Mp$\rm c^{-1}$. Precisely determining $\Omega_{\rm B}h^2$ by other means can narrow down the range for the calculated primordial abundances, and thus confirm whether the observationally inferred primordial abundances are consistent.

The Wilkinson Microwave Anisotropy Probe (WMAP) achieved high precision in measuring $\Omega_{\rm B}h^{2}$. Using measurements of the cosmic microwave background (CMB) radiation, the value $\Omega_{\rm B}h^{2}= 0.0227\pm 0.0006$ was inferred (\cite{Dunkley08}). This value is in excellent agreement with that derived from BBN via the measured \element[][2]{H}/\element[][1]{H}  and  \element[][4]{He}/\element[][1]{H} ratios, although the systematic uncertainties in the latter
make \element[][4]{He} a poor discriminant. 
However, for other primordial isotopes the results are not in such good agreement (\cite{CocVangioni05}, their Fig. 4). The baryon densities inferred from \element[][3]{He} and \element[][7]{Li} abundances differ significantly. \element[][7]{Li} has the greatest deviation. The deviation of the observationally inferred primordial \element[][7]{Li} abundance from that deduced from WMAP and from \element[][2]{H} via BBN has become known as the ``lithium problem''.

The \element[][3]{He} discrepancy can be accommodated within large uncertainties concerning the mechanisms in its production and destruction in stars (\cite{VangioniFlam03}).
 It is not clear that the \element[][7]{Li} discrepancy can be explained as easily. 

The primordial Li abundance was first deduced from observations of halo stars by \cite{spitex2} when they discovered the \element[][7]{Li} plateau, now known as the Spite plateau, with a mean abundance of \textit{n}(\element[][7]{Li})/\textit{n}(\element[][]{H}) = $1.12 (\pm0.38)\times10^{-10}$. Many new obsevationally determined abundances have been published, the values of which can differ significantly, for example from the low value of \textit{n}(\element[][7]{Li})/\textit{n}(\element[][]{H}) = $(1.23^{+0.68}_{-0.32}\times10^{-10}) (95{\%}$ confidence limits) (\cite{Ryan00}), to a 	decidedly larger value of  \textit{n}(\element[][7]{Li})/\textit{n}(\element[][]{H}) = $2.34\times10^{-10}$ (Mel\'{e}ndez {\&} Ram\'{i}rez 2004). In comparison the abundance calculated from WMAP and from \element[][2]{H} via BBN is \textit{n}(\element[][7]{Li})/\textit{n}(\element[][]{H}) = $(4.15^{+0.49}_{-0.45}\times10^{-10})$ (\cite{Cocetal04})
or (\element[][7]{Li})/\textit{n}(\element[][]{H}) = $(4.26^{+0.73}_{-0.60} \times 10^{-10})$ (\cite{cfo}, \cite{cyburt}) . It is obvious that the theoretical and observational values are not in accord with each other. The observed \element[][7]{Li} abundance is a factor of 2 to 3 times lower than that given by BBN/WMAP, far beyond the stated range of systematic uncertainties from observational analysis or BBN calculations.

Several possibilities have emerged in an attempt to explain this \element[][7]{Li} discrepancy, but none of these, as yet, fully explains the problem. Broadly speaking these lie in the categories:
\begin{enumerate}
	\item The discrepancy is due to unrecognised or underestimated systematic errors in the calculations of the observationally inferred present-day Li abundance.
	\item The stars studied have destroyed some of the Li with which they were formed.
	\item Some Galactic Li was destroyed before these stars formed.
	\item The discrepancy is due to systematic uncertainties in the nuclear cross sections used in BBN calculation.
	\item Standard BBN has failed to accurately predict the primordial abundances.
\end{enumerate}

A possible solution in the second category is diffusion. Several studies have been conducted on this subject recently, e.g. \cite{RMR05} and \cite{Kornetal06}. There it is suggested that the primordial Li is depleted in the star by diffusion, through gravitational settling, to layers in which the Li can not be detected. The models used in this process are manipulated to minimize destruction of Li due to nuclear burning. However, they do not completely manage to eliminate this destruction. Although these predictions do lead to a possible solution to the Li problem, they are still subject to questions and uncertainties.  Using atomic diffusion alone does not recreate the plateau, but causes a drop in Li abundance in stars with $\textit{T}_{\rm eff}$ $>$ 6000 K. It is also often hard to explain the smallness of the star to star scatter in \element[][7]{Li} using these processes. To solve these problems some form of turbulence at the bottom of the convective zone has to be invoked. The source of this turbulence is still uncertain, as is the amount needed, which can vary from star to star. The turbulence is also dependent on the temperature scale used. If a cooler $\textit{T}_{\rm eff}$ scale is used then more turbulence is needed. However, observations of \element[][6]{Li} in the atmospheres of metal-poor turn off stars (Asplund et al. 2006, \cite{GarciaPerezetal08}) rule out the use of a more turbulent scheme, as it would destroy the \element[][6]{Li}. It would therefore seem that the success of this solution to the lithium problem rests on an accurate $\textit{T}_{\rm eff}$ scale, and confirmation of whether \element[][6]{Li} is present on the surface of metal-poor stars.  However, it should be noted that the presence of \element[][6]{Li} is still in question. It has been shown (\cite{Cayreletal07} and \cite{GarciaPerezetal08}) that a reappraisal of previously derived \element[][6]{Li} abundance should be undertaken. The presence of \element[][6]{Li} also causes problems if \element[][7]{Li} undergoes nuclear burning. Although the models of diffusion and turbulence are tuned so as to reduce this burning it is not clearly stated whether they eliminate it completely. As \element[][6]{Li} is destroyed at a lower temperature than \element[][7]{Li}, the presence of \element[][6]{Li} would rule out the possibility of any models in which \element[][7]{Li} is destroyed.

Possible explanations in the third category are relatively new. One such
mechanism involving stellar processing is that of \cite{Piauetal06}.
However, the amount of material which this requires to be cycled through
stars creates other problems; the implications of such models require
further investigation.

It is of course possible that certain nuclear cross sections used in BBN calculations have been poorly determined.  The effect of changing the yields of certain BBN reactions was
recently considered by \cite{Cocetal04}.  They found for example,
that an increase of the \element[][7]{Be}(d,p)2\element[][4]{He} reaction by a factor of 100
would reduce the \element[][7]{Li} abundance by a factor of about 3 in the WMAP $\eta$ range. 
This reaction has
since been remeasured and precludes this solution (\cite{Ang05}). 
There is also the  possibility
that systematic errors in the
\element[][3]{He}$(\alpha,\gamma)$\element[][7]{Be} reaction are the cause of the \element[][7]{Li}
discrepancy. This channel was considered in detail in Cyburt et al. (2004).  
Although the absolute value of the cross section
for this key reaction is known relatively poorly both experimentally
and theoretically, the agreement between the standard solar model and
solar neutrino data provides additional constraints on variations
in this cross section.  Using the standard solar model of
\cite{bah}, and recent solar neutrino data (\cite{sno}), one can exclude systematic
variations of the magnitude needed to resolve the BBN \element[][7]{Li}
problem at the $\ga 95\%$ confidence level (\cite{cfo4}).

The fifth category represents an interesting challenge, with several cosmological and particle physics possibilities arising which could affect BBN (\cite{CocVangioni05}). For example,  the variation of the fine structure constant can induce a variation in the deuterium binding energy and could
yield a decrease in the predicted abundance of \element[][7]{Li} (\cite{dfw}, \cite{cnouv}).  The modification of the expansion rate during BBN (\cite{Salati03}) will also affect the light element abundances. Another possibility is that gravity is not described by general relativity but is instead attracted toward general relativity during the evolution of the Universe (\cite{dn1}, \cite{dn2}). 
BBN has been extensively studied in that scenario (\cite{dp}, \cite{couv}). Finally, it may be 
that physics beyond the standard model is responsible for the post BBN processing of the light elements. 
One possibility recently discussed is that
particle decay after BBN could lower the \element[][7]{Li} abundance and produce some \element[][6]{Li} as well (\cite{Jedamzik04}). This has been investigated in the framework of the constrained minimal
supersymmetric standard model if the lightest supersymmetric particle is assumed to be the
gravitino (\cite{feng}, \cite{eov}, \cite{Hamaguchietal07}). Some models have been found which accomplish these goals (\cite{Jedamzik06}, \cite{cefos}).
 However, all of these possibilities lack confirmation at present. Also the other four categories of explanation listed above need to be ruled out in order for the last to gain significant favour.

It is the first category of explanation that we address in this work. The largest of the uncertainties arises from the uncertain effective temperature scales for metal-poor stars, which we now examine in greater detail, before going on to calculate new effective temperatures and lithium abundances for a sample of metal-poor main-sequence stars, which is the ultimate aim of this work.

\subsection{Effective Temperature ($T_{\rm eff}$) Scale Problems}
\label{sec:EffectiveTemperatureTEffScaleProblems}

In calculating the abundances of Li, $A$(Li)\footnote{$A$(Li)$\equiv log_{10} \left(\frac{n(\rm Li)}{n(\rm H)}\right)+12.00$}, from spectral measurements, the effective temperature is the most important atmospheric parameter, as $A$(Li) has a high sensitivity to temperature: $\partial{A}/\partial{T_{\rm eff}} {\sim}$ 0.065 dex per 100 K for halo main-sequence turnoff stars. The temperature scale adopted by different authors varies considerably. A comparison of effective temperatures adopted by \cite{Ryanetal01} and \cite{RamirezMelendez05} shows differences for very low metallicities ([Fe/H] $<$ -3) by as much as 500 K, with typical differences of $\sim200$ K. Concerning the very metal-poor star G64-37 which features in the present paper, the difference in temperature between \cite{Ryanetal99}, at 6240 K, and \cite{MelendezRamirez04}, at 6775 K, is 535 K. In these cases \cite{RamirezMelendez05} and \cite{MelendezRamirez04} have the hotter temperature scale, and hence they infer higher primordial abundances as noted above. 

Several methods are routinely adopted in calculating the effective temperature. These include using the strong Balmer line wings (e.g. Asplund et al. 2006), photometric methods including the Infrared Flux Method (IRFM) (e.g. Ram\'{i}rez \& Mel\'{e}ndez 2005) and a spectroscopic method that utilizes the temperature dependence of the atomic-level populations, as given in local thermodynamic equilibrium  (LTE) by the Boltzmann equation.

In the first method, the Balmer line wings are used as a temperature indicator as they have strong temperature sensitivities, whilst maintaining a low sensitivity to other physical parameters, i.e. log $\textit{g}$ (\cite{Fuhrmannetal92}). $H\alpha$  in particular has a high sensitivity to $T_{\rm eff}$, $\sim$ 0.5 \AA\ in the width of the wings per 100 K, and is the least sensitive to other parameters. This technique is widely used, e.g. \cite{Fuhrmannetal94}, \cite{Barklemetal02}, and \cite{Asplundetal06}. It assumes that the wings are formed in LTE. This assumption makes it simpler to model, thus leading to more sturdy temperature scales.

The preferred temperature scale in many studies is that derived by the IRFM. This method uses IR photometry and model fluxes to constrain the temperature. The basic idea is to compare a theoretically calculated infrared flux with an observed infrared flux to derive an angular diameter, given by the equation:
\begin{equation}
	\theta=2\sqrt{\frac{{\textit{f}}_{\rm obs,IR}}{{\textit{f}}_{\rm theo,IR}}} \,,
\end{equation}
where ${\textit{f}}_{\rm obs,IR}$ and ${\textit{f}}_{\rm theo,IR}$ are the observed and theoretical infrared fluxes respectively. This $\theta$ can then be used, along with the integrated observed flux, to calculate an effective temperature, using the equation:
\begin{equation}
	\textit{T}_{\rm eff}=\left(\frac{4{\textit{F}}_{\rm obs}}{\sigma\theta^{2}}\right)^{\frac{1}{4}} \,,
\end{equation}
where ${\textit{F}}_{\rm obs}$ is the observed integrated flux at the Earth and $\sigma$ is the Stefan-Boltzmann constant. This is an iterative process where the ${\textit{T}}_{\rm eff}$ is used to refine the theoretical fluxes, thus refining the angular diameter and the effective temperature. In essence the method comes down to simultaneously solving equations 1 and 2 to derive values of $\theta$ and $\textit{T}_{\rm eff}$. Due to the lower sensitivity of the infrared flux to temperature compared to shorter wavelengths,  when equation 1 and equation 2 are solved simultaneously the calculated temperature for the star is better defined, and hence more accurate (see \cite{Blackwelletal79} Fig.1). Mean random errors for recent work using this method are $\sim$ 60 - 75 K, depending on the type of star analysed, from \cite{RamirezMelendez05}, and $\sim$ 70 K from \cite{Alonsoetal99}. Temperatures from this method have been used to calibrate other photometric indices (e.g. \cite{Magain87} and Alonso et al. (1996))

The third method usually relies on the assumption of LTE. This implies that the Boltzmann equation can be utilized in the derivation of the effective temperatures. Boltzmann's equation contains an exponential term in $\chi_{i}/T$, where $\chi_{i}$ is the excitation energy of the given energy level $\textit{i}$, and $\textit{T}$ is the temperature of the gas. Each measured equivalent width (EW) of an element, in most cases \ion{Fe}{I}, is used to determine an abundance for that element. If the temperature used in the calculation is too high, then an over-population of the more excited levels will be calculated with the Boltzmann equation. This will lead to the inference that there are already more absorbers in higher levels, and hence it would then be calculated that fewer absorbers were needed to reproduce the measured EW. A dependence of abundance on excitation energy would then be seen due to these miscalculated populations. The correct temperature will therefore be the one that nulls any trend in the plot of abundance against excitation energy (e.g. \cite{PetersonCarney79}).

However, all methods have, along with their strengths, significant deficiencies. \cite{Barklem07} has questioned whether the assumption that the wings of the Balmer lines form under conditions of LTE is reliable. This therefore casts doubt on the accuracy of the derived temperatures, with a possible temperature rise of 100 K compared to those derived in LTE. Along with this there are also uncertainties in the way hydrogen-hydrogen collisions, so-called self-broadening, are treated. \cite{Barklemetal00} found differences in $T_{\rm eff}$ of up to 150 K between the their theory and the Ali-Griem treatment of self-broadening. This difference was also found by \cite{Bonifacioetal07} using a Ali-Griem theory which has modified broadening coefficients, which allow for a good match to solar Balmer lines. The IRFM relies heavily on photometry and modelled fluxes to derive the temperatures. It can therefore suffer from photometric errors. This method also has a dependency on model atmosphere. Finally, temperatures derived using the excitation energy technique currently depend on the accuracy of assumptions about LTE, about the structures of 1D model atmospheres and the  errors in $gf$ values, damping values and equivalent width measurements. It has been shown (\cite{Theveninetal99} and \cite{Asplundetal99}) that the assumptions of 1D, LTE atmospheres are not always suitable for modelling radiative transfer of Fe lines in metal--poor stars, although \cite{Grattonetal99} infer there is little sensitivity to non-LTE. It is thus unclear how strong these effects are.

In previous work by some of the present authors (Ryan et al. 1999), temperatures were derived using a calibration of BVRI colours and medium resolution ($\Delta\lambda$ $\approx$ 1\AA) spectroscopy, tied principally to the Magain (1987) $\textit{b-y}$ calibration of IRFM temperatures. In Ryan et al. (2000), an adjustment of +120 K was made to the temperature scale to make it similar to the IRFM scale of \cite{Alonsoetal96}. \cite{Alonsoetal96} report similarly a mean difference of 112 K between their IRFM values and Magain's. The aim of the present work was to explore the use of the excitation energy technique to further constrain the effective temperature scale. In other studies where  excitation-energy temperatures have been derived, only limited attention has been paid to quantifying the random and systematic uncertainties on those values. Given the importance of the temperature scale to the lithium problem, we endeavor here to track the uncertainties more closely.

\subsection{1D-LTE vs. 3D/1D-Non-LTE}
\label{sec:LTEVsNonLTE}

Because the calculation of an effective temperature from the excitation-energy technique requires that the populations of the different energy levels be known, it is necessary to investigate non-LTE effects in metal-poor stars and their effect on the temperature scale calculation.

As discussed by Asplund et al. (1999) the level populations for the Fe lines are affected by departures from LTE, as radiation fields higher than the local Planck function lead to over-ionisation of the atom compared to the LTE case. That is, applying Saha's equation with the local value of the temperature at some depth in the atmosphere will underestimate the degree of ionisation. Moreover, different energy levels will be affected differently by non-LTE, so it is not possible to apply a single ionisation correction to all levels of a given ionisation state. A second factor that could affect excitation equilibrium is that of the use of 3D atmospheres. It can be seen in \cite{Asplund05} (his Figure 8) that there can be a significant $\chi$-dependent effect on the abundances produced. However, in the same paper it is stated that non-LTE corrections appear to have the opposite effect to the 3D corrections and these corrections also become larger when 3D atmospheres are used. This could lead to 3D non-LTE becoming very similar to 1D non-LTE. It is not known, however, that this is always the case. Non-LTE effects, 3D effects, and their combination relative to 1D-LTE will therefore have to be studied to assess the exact impact on the temperatures derived by this technique. 

However, it is well known that non-LTE calculations for \ion{Fe}{I} in metal-poor stars are quite uncertain, principally because of uncertainty in the role of collisions with hydrogen. Th\'{e}venin \& Idiart (1999) and \cite{Grattonetal99} reached opposite conclusions about the importance, or otherwise, of non-LTE in such stars. 3D calculations are also relatively rare to date. Because of these uncertainties, we proceed in two stages.  For the present paper we shall concentrate on a 1D LTE analysis, and this will be the focus for the remainder of this paper. In a second paper we will examine the complexities introduced by attempting a non-LTE analysis.


\section{Observations}

\subsection{Sample Selection}
\label{sec:SampleSelection}

The sample of stars was taken from one constructed by Ryan et al. (1999) which contains 24 metal-poor stars spread around the sky. It is a carefully selected sample at the Population II main-sequence turnoff, with as few star-to-star variations as possible. This included limiting the effective temperature to a range of 6100 K $\la \textit{T}_{\rm eff} \la$ 6300 K and the metallicity to -3.5 $\la$ [Fe/H] $\la$ -2.5. These restrictions allow the sample to have roughly the same evolutionary state, in this case around the MS turn-off, which also has the effect of limiting log $\textit{g}$. The low metallicity restrictions mean that we are sampling stars that contain material that has undergone a limited amount of processing since the Big Bang. The sample was previously studied to determine not only Li abundances, but also those of magnesium (\cite{Arnoneetal05}), based on red spectra.

The intention in the present work was to obtain blue high--resolution spectra for a subset of these stars. This was to enable us to use the excitation-energy technique to determine the $\textit{T}_{\rm eff}$, and determine what, if any, correction is required to the photometric $\textit{T}_{\rm eff}$ scale used previously. This also allows for a comparison between the excitation-energy $\textit{T}_{\rm eff}$ and those scales used by \cite{Asplundetal05} and \cite{MelendezRamirez04}.

\subsection{Data Acquisition and Reduction}
\label{sec:DataAcquisitionAndReduction}

Spectra of 16 of the program stars plus two standards, out of the 22 program stars and two standards, HD140283 and HD74000, in the \cite{Ryanetal99} study, were obtained using the UCLES instrument on the 3.9m Anglo-Australian Telescope (AAT) during two separate observing runs. The first run, July 31-August 3 2006, was undertaken by S.G. Ryan and J.D. Tanner. The second run, March 9-11 2007, was undertaken by A. Hosford and A.E. Garc\'{i}a Per\'{e}z. The setup was the same for both observing runs: the 79 lines/mm grating was used, centred at 4174 \AA\ with a slit width of 1.0\arcsec, which achieves a resolving power of $R\equiv\lambda/\Delta\lambda=43000$. The wavelength range is from 3700 \AA\ - 4900 \AA. In Table \ref{table:1} we present the basic information including the photometric (and partially spectroscopic) temperature, $\textit{T}_{\rm phot}$ (Ryan et al. 1999) for each of the stars, along with the S/N per 0.025~\AA\ pixel near 4300 \AA\ from the present work.

The data reduction was done using standard routines within the IRAF package.

\begin{table*}[!t]
	\caption{Background infomation for the stars analysed in this work.}
	\label{table:1}
		\centering
		\begin{tabular} {l c r c c c c c}
		\hline\hline
Star name	&	RA (1950)	&	DEC (1950)	&	V	&	B-V	&	S/N @ 4300Å	&	$\textit{T}_{\rm phot}$ (K)	&	Initial [Fe/H]	\\
    \hline
BD-13$^{\circ}$3442	&	11h 44m 18s	&	$-13^{\circ} 49' 54''$	&	10.37	&	0.33	&	80	&	6210	&	$-2.73$	\\
BD+20$^{\circ}$2030	&	08h 13m 13s	&	$19^{\circ} 51' 24''$	&	11.20	&	0.40	&	95	&	6200	&	$-2.64$	\\
BD+24$^{\circ}$1676	&	07h 27m 39s	&	$24^{\circ} 11' 42''$	&	10.80	&	0.40	&	100	&	6170	&	$-2.38$	\\
BD+26$^{\circ}$2621	&	14h 52m 00s	&	$25^{\circ} 46' 12''$	&	11.01	&	0.50	&	95	&	6150	&	$-2.54$	\\
BD+26$^{\circ}$3578	&	19h 30m 29s	&	$26^{\circ} 17' 06''$	&	9.37	&	0.43	&	110	&	6150	&	$-2.24$	\\
BD+3$^{\circ}$740	&	04h 58m 38s	&	$04^{\circ} 02' 24''$	&	9.82	&	0.38	&	100	&	6240	&	$-2.70$	\\
BD+9$^{\circ}$2190	&	09h 26m 35s	&	$08^{\circ} 51' 24''$	&	11.14	&	0.36	&	95	&	6250	&	$-2.83$	\\
CD-24$^{\circ}$17504	&	23h 04m 39s	&	$-24^{\circ} 08' 42''$	&	12.18	&	0.32	&	100	&	6070	&	$-3.24$	\\
CD-33$^{\circ}$1173	&	03h 17m 34s	&	$-33^{\circ} 01' 21''$	&	10.91	&	0.39	&	100	&	6250	&	$-2.91$	\\
CD-35$^{\circ}$14849	&	21h 30m 48s	&	$-35^{\circ} 39' 12''$	&	10.63	&	0.37	&	105	&	6060	&	$-2.38$	\\
CD-71$^{\circ}$1234	&	16h 02m 18s	&	$-71^{\circ} 14' 00''$	&	10.44	&	0.46	&	95	&	6190	&	$-2.60$	\\
G64-12	&	13h 37m 30s	&	$00^{\circ} 12' 54''$	&	11.49	&	0.41	&	110	&	6220	&	$-3.24$	\\
G64-37	&	13h 59m 53s	&	$-05^{\circ} 24' 18''$	&	11.14	&	0.36	&	100	&	6240	&	$-3.15$	\\
LP635-14	&	20h 24m 13s	&	$-00^{\circ} 47' 00''$	&	11.33	&	0.47	&	115	&	6270	&	$-2.66$	\\
LP815-43	&	20h 35m 21s	&	$-20^{\circ} 36' 30''$	&	10.91	&	0.39	&	95	&	6340	&	$-3.00$	\\
HD 84937	&	09h 46m 17s	&	$13^{\circ} 59' 18''$	&	8.28	&	0.42	&	110	&	6160	&	$-2.12$	\\
\hline\hline
HD 140283	&	15h 40m 22s	&	$-10^{\circ} 46' 18''$	&	7.24	&	0.46	&	170	&	5750	&	$-2.54$	\\
HD 74000	&	03h 38m 31s	&	$-16^{\circ} 09' 36''$	&	9.671	&	0.43	&	150	&	6040	&	$-2.02$	\\
\hline

\end{tabular}
\end{table*}

\section{Data Analysis}
\label{sec:DataAnalysis}

Analysis of the data proceeded in, broadly, two stages. Firstly, the equivalent widths of spectral lines of Fe were measured. For the second stage we use WIDTH6 (Kurucz \& Furenlid 1978) to calculate the abundances for each line, from which the physical parameters can be deduced. The remainder of this section discusses these stages.

\subsection{Equivalent Width Measurement and Data Quality Checks}
\label{sec:EquivalentWidthMeasurementAndDataQualityChecks}

\begin{figure}
\centering

	\psfig{file=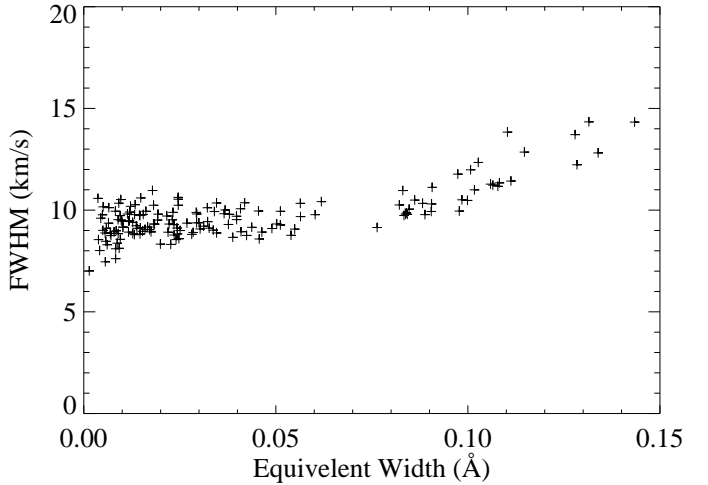,width=9cm}
	\caption{Plot of $\textit{v}_{\rm FWHM}$ versus equivalent width for the retained lines in the star HD140283}
\label{fig:HD140283FWHMplot}
\end{figure}

Equivalent widths (EW) of apparently unblended \ion{Fe}{I} and \ion{Fe}{II} lines in the spectrum of each star were measured by fitting Gaussian profiles using the IRAF `splot' program. Graphs of the Gaussian FWHM in velocity space ($\textit{v}_{\rm FWHM}$) were plotted against the measured equivalent width of the line, see Figure (\ref{fig:HD140283FWHMplot}). Due to the nature of echelle spectra, in which $\Delta\lambda/\lambda $ is constant throughout the spectrum (where $\Delta\lambda$ is the width of a pixel in wavelength units), it is possible to use these plots as checks of the quality of the measurements. The weak lines, EW $\la$ 90 m\AA, form a plateau due to the constancy of the FWHM of the Doppler core of lines on the linear part of the curve of growth. The Doppler core is of course broadened by the instrumental profile, which likewise has a constant width in velocity units. For intermediate strength and stronger lines, the plot curves upward due to the increase in their line wing strength, leading to larger FWHM. Lines whose equivalent widths fell significantly outside of these trends were re-measured or discarded as blends or noise-dominated measurements, and were rejected from further calculations. The equivalent widths measured in each star are listed in Table A1, A2 and A3 in the appendix to this paper. (The appendix is available online only.)

A comparison between equivalent widths of the Fe lines in the star CD-24$^{\circ}$17504 measured by \cite{Norrisetal01} and remeasured by A. Hosford showed good precision. The standard deviation of the difference between the two sets of equivalent widths was 1.3 m\AA\ with a mean of $-0.58$ m\AA\ for lines measured from the same spectra. Lines which were measured in two spectra, one obtained from \cite{Norrisetal01} at S/N $\approx$ 100, and one obtained by A. Hosford at S/N $\approx$ 50, were also compared. This comparison gave a standard deviation of 5.8~m\AA\ and a mean of $-0.39$ m\AA. To increase the accuracy of the EW  measurements for the star CD-24$^{\circ}$17504 we used a combination of EW measurements from both the \cite{Norrisetal01} spectra and the A. Hosford spectrum, weighting each measurement by the inverse of the uncertainty in the EW ($\sigma_{\rm EW}$). This $\sigma_{\rm EW}$ is calculated using the equation:
\begin{equation}
\sigma_{\rm EW}=\frac{\Delta\lambda_{pix}\sqrt{M}}{S/N} \,,
\end{equation}
where $\Delta\lambda_{pix}$ is the width of a pixel in wavelength units, $M$ is the number of pixels comprising the full width of the spectral line and $S/N$ is the signal-to-noise of the spectra.

\subsection{Abundance Calculations}
\label{sec:AbundanceCalculations}

To calculate the abundances from the measured lines we use the single line radiative transfer program WIDTH6. This program  calculates the EW of an unblended absorption line for a 1D model atmosphere. It varies the starting abundance, iteratively calculating an EW until that equals the measured EW for the line.

The KURUCZ06 (http://kurucz.harvard.edu/grids.html, downloaded in Nov 2006) model atmospheres have been adopted for this work as it has been stated (\cite{Castellietal97}) that these no-overshooting models are preferable in deriving accurate temperatures to earlier Kurucz overshooting models (\cite{Kurucz1993}). This is due to the effect the different methods of treating convection have on the temperature structure of the atmosphere, and therefore on the spectral features. The no-overshooting models lead to a better fit to observable spectral features at different temperatures. These models are interpolated in $\textit{T}_{\rm eff}$, log $\textit{g}$ and [Fe/H] over the ranges $3500 \leq \textit{T}_{\rm eff} \leq 50000$, $0.0 \leq$ log $\textit{g} \leq 5.0$ and $-0.5 \geq [Fe/H] \geq -4.0$. A comparison between temperatures derived using the different overshooting models can be seen in Sect. 3.3.

The log \textit{gf} values of the \ion{Fe}{I} lines were compiled from several sources: \cite{Blackwell1etal79}, \cite{Blackwell2etal79}, \cite{Blackwelletal80}, \cite{Blackwelletal82}, \cite{Blackwell2etal82}, collectively the `Oxford \ion{Fe}{I} consortium', \cite{Fuhretal88}, \cite{Bardetal91}, \cite{O'Brianetal91}, \cite{BardKock94} and \cite{Theveninetal89}; in some cases the mean of several log \textit{gf} values is adopted, see Table A1 for details. The \ion{Fe}{II} log \textit{gf} values are compiled from: \cite{Hannafordetal92}, \cite{HeiseKock90}, \cite{KrollKock87}, \cite{Moity83} and \cite{SchnabelKockHolweger99}. Values given by \cite{Moity83} are corrected by $-0.06$ dex, for those lines with upper-level energy $<$ 48000 cm$^{-1}$, and $-0.24$ dex for lines with upper-level energy greater than the above value; this follows the suggestion by \cite{Fuhretal88}. Damping values have been calculated using the tabulations of \cite{AnsteeO'Mara95}, \cite{BarklemO'Mara97}, and \cite{Barklemetal98}. \cite{Ryan98} noted that the damping treatment has a significant effect on the excitation energy temperatures because of the $\chi$-dependence of damping treatments.

\subsection{Adopted Approach to Constraining Parameters}
\label{sec:AdoptedApproachToConstrainingParameters}

Several other physical parameters of a star, in addition to temperature, could in principle be constrained using WIDTH6, given appropriate data. These are log \textit{g}, metallicity and microturbulence. Metallicity is determined from the calculated Fe abundance. Log \textit{g} could be constrained by ensuring that the abundances derived for \ion{Fe}{I} and \ion{Fe}{II} lines are equal, and microturbulence is found by making sure there is no trend between the measured equivalent widths of the lines and the abundance.

It would, in principle, seem possible to simultaneously constrain both log \textit{g} and $\textit{T}_{\rm eff}$. However, as previously mentioned, there are concerns about the non-validity of the assumption of LTE for \ion{Fe}{I} lines in metal-poor turnoff stars. This would therefore make it dangerous to try to constrain $\textit{T}_{\rm eff}$ at the same time as log \textit{g}. This concern became evident when $\textit{T}_{\rm eff}$ and log \textit{g} where simultaneously constrained via a WIDTH6 analysis for the star HD140283, using \ion{Fe}{I} and \ion{Fe}{II} lines. For this star results converged at $\textit{T}_{\rm eff}/\rm log \textit{g}/[Fe/H]/\xi = 5573/3.1/-2.67/1.5$ using the KURUCZ06 models\footnote{We also tried this with the KURUCZ93 models and found 5439/2.8/-2.79/1.5.}. However, we know that the Hipparcos parallax gives a reliable value for log \textit{g} for HD140283: log $\textit{g} = 3.73 \pm 0.11$ for a mass of 0.8360 $M_{\sun}$ as taken from the Yonsei-Yale isochrones (http://csaweb.yonsei.ac.kr/$\sim$kim/yyiso.html). The uncertainty is dominated by the parallax error. Clearly the log $\textit{g}$ value we have derived is too low. This star has also appeared in many other studies leading to temperatures around 5750 K (Table 2), approximately 200 K hotter than we derived. It can then be seen that for this star, attempting to constrain $\textit{T}_{\rm eff}$ and log $\textit{g}$ simultaneously from \ion{Fe}{i} and \ion{Fe}{ii} lines drives the gravity and possibly the temperature values down to unusually low levels. Non-LTE effects may be responsible. It is, however, still viable to use WIDTH6 to constrain the $\textit{T}_{\rm eff}$. This is because the main non-LTE correction involved in Fe is expected to be overionisation, which may significantly affect the abundance of \ion{Fe}{I} and therefore have a substantial effect on log $g$ when calculated through ionisation balance. The correction on $\chi$ is very roughly the same for all levels within a particular star. This can be seen from Figures 2 and 3 from \cite{Colletetal05}, while \cite{Asplund05} also states this view. \cite{Shchukinaetal05} do find that the effect can be different for different excitation potentials, however, this seems far more pronounced when using 3D atmospheres, compared to using 1D atmospheres. Therefore overionisation will have less affect on the trend between [Fe/H] and $\chi$, and it is still acceptable to use this to constrain $\textit{T}_{\rm eff}$,  subject to the caveats in Sect. 1.3.

\begin{table}[htbp]
\begin{minipage}[t]{\columnwidth}
\caption{List of results for log \textit{g} and $\textit{T}_{\rm eff}$ of the star HD140283 derived using different techniques.}
	\label{table:2}
	\centering
	\renewcommand{\footnoterule}{}  
		\begin{tabular} {l l l}
		\hline\hline
		Study & Temperature & $\textit{T}_{\rm eff}/\rm log \textit{g}$ \\
		\hline
		This work & Simultaneous fit & $5573/3.1$ \footnote{The uncertainties are 75 K and 0.15 dex respectively} \\
		\cite{Asplundetal06} & Balmer line wing & $5753/3.7$ \footnote{The uncertainties are 30 K and 0.04 dex respectively} \\
		\cite{Ryanetal96} & Photometry & $5750/3.4$ \\
		\cite{Alonsoetal96} & IRFM & $5691/4.0$ \footnote{The uncertainties are 69 K and 0.5 dex respectively}\\
		\hline
			
		\end{tabular}
		\end{minipage}
\end{table}

It is therefore preferable to constrain log \textit{g} independently of deducing $\textit{T}_{\rm eff}$, rather than using Fe lines to constrain both. There are several options for doing this. In rare cases the Hipparcos parallax could be used. However a large proportion of the program stars do not have these data available, and for the stars where it is available the parallax errors are large. Alternatively, WIDTH6 could be used with a fixed temperature, such as an assumed photometric temperature, $\textit{T}_{\rm phot}$ from previous work, to constrain log \textit{g} through ionisation balance. This method still leads to concerns with the non-validity of LTE and the sensitivity of the method to the assumed temperature. As a third option, theoretical isochrones for old, metal-poor stars near the main sequence turnoff can be used. This method, however, leads to a range of possible values for log \textit{g}, mainly due to the uncertainty in age and evolutionary state, i.e. whether a given star is pre- or post-turnoff.

Because of the various concerns raised above, we adopted the following two stage procedure to analyse the stars. The first step is to use the Hipparcos parallax of the star if available. Table \ref{table:3} contains gravities calculated from the Hipparcos parallax, log $\textit{g}_{\rm Hipp}$, for each star, where possible, with masses taken from the Yonsei-Yale isochrones with an assumed age of 12 Gyr and a metallicity suitable for the star. We obtain two Hipparcos gravities due to the different stellar masses the star could have depending on whether it is on the main sequence (MS) or subgiant branch (SGB). In practice, the difference between these two values is small compared to the uncertainty arising in the parallax, and so we quote at most one value of log $\textit{g}_{\rm Hipp}$ for each star. Many of the stars' parallaxes have extremely large errors, to the point where they would have been unsatisfactory to use, and are excluded. In the case of the HD stars, we have taken the log $\textit{g}_{\rm Hipp}$ as the final gravity as they have small errors. However, for the other five stars with log $\textit{g}_{\rm Hipp}$ in Table \ref{table:3}, we have taken log $\textit{g}_{\rm Hipp}$ to be only an interim value. A theoretical gravity, log $\textit{g}_{\rm iso}$, is then used to determine a temperature as described below. We have used$\textit{T}_{\rm phot}$ to calculate the Hipparcos gravity, log $\textit{g}_{\rm Hipp}$, but note that it has only a low sensitivity to $T_{\rm phot}$ of +0.04 dex per 100 K.

We also run WIDTH6 with a fixed $\textit{T}_{\rm phot}$ (see Table 1) to find an interim (ionization-balance) log \textit{g}, which we call log $\textit{g}_{\rm Tphot}$. The results of this analysis are found in Table~ \ref{table:3}. The sensitivity of log $\textit{g}_{\rm Tphot}$ to $T_{\rm phot}$ is +0.15~ dex per 100~ K. The difference between the gravities, log $\textit{g}_{\rm Tphot}$ - log $\textit{g}_{\rm Hipp}$, ranges from -0.18 to -0.52, with a mean of -0.37. This suggests that this method, using the $\textit{T}_{\rm phot}$ and the Fe ionisation balance, underestimates the gravity by $\sim$0.4~ dex. This could be due to $T_{\rm phot}$ being too cool by $\sim$270~ K, or may be evidence of non-LTE effects (overionisation) affecting \ion{Fe}{I}. We make no further use of the log $\textit{g}_{\rm Tphot}$.

\begin{figure}
\centering
\vbox{
	\psfig{file=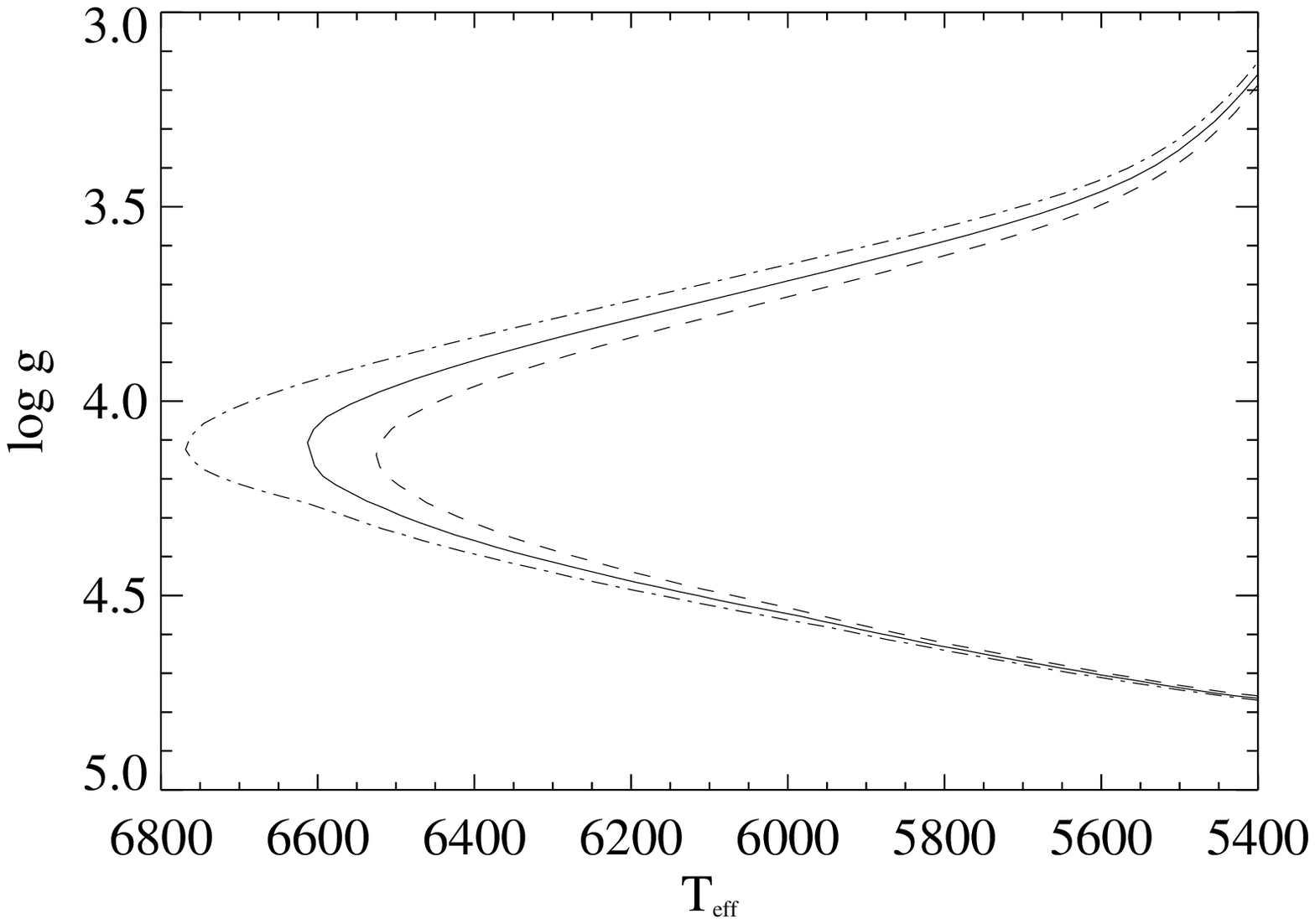,width=9cm}}
\vbox{	
	\psfig{file=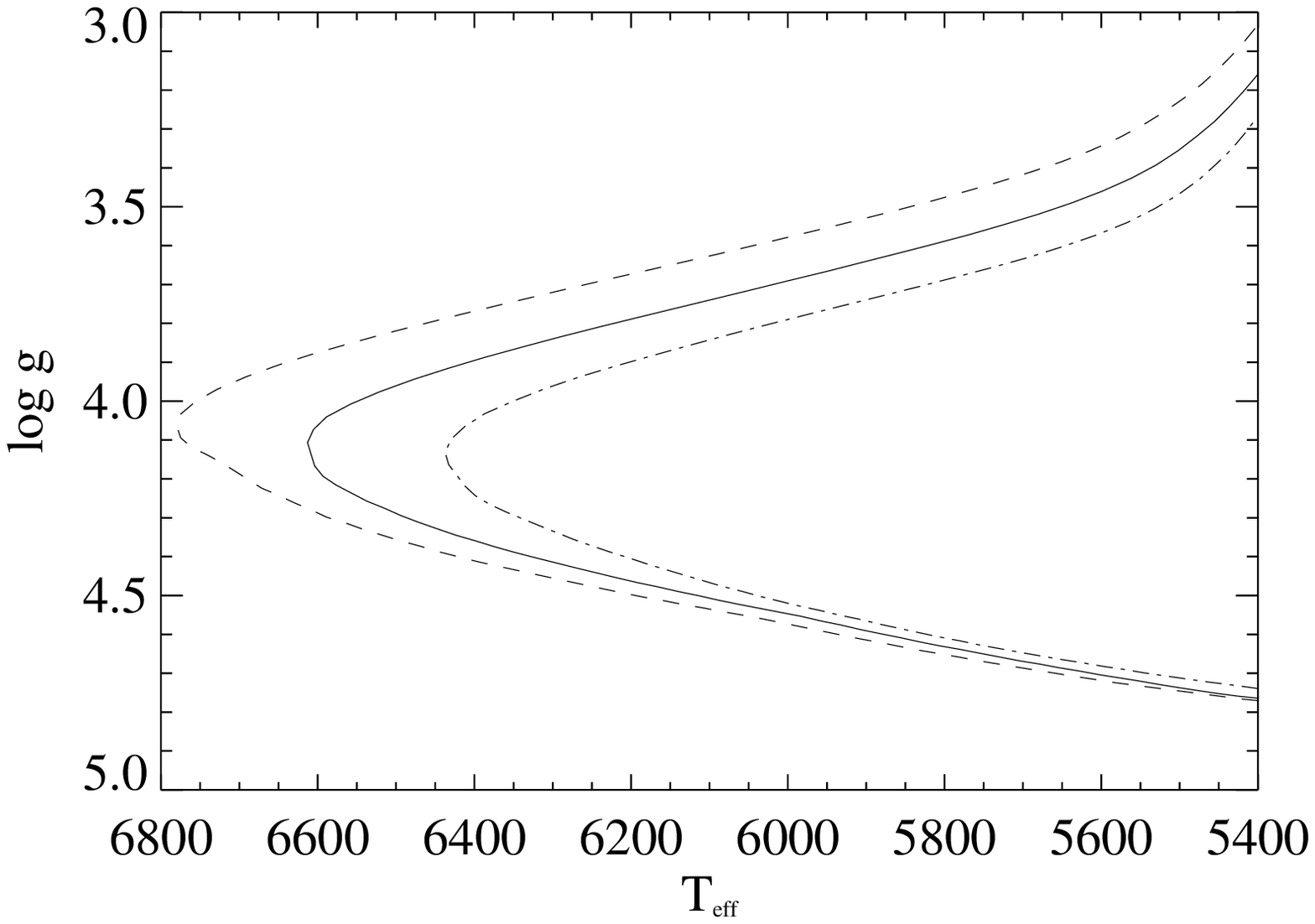,width=9cm}}
	\caption{Theoretical isochrones used in the determination of log $\textit{g}$. Different ages are represented in the top panel,with the dashed line 13 Gyr, the solid line 12 Gyr, and the dot dashed line 11 Gyr. The bottom panel gives examples for different [Fe/H]. The dashed line is -3.3, the solid line -2.5 and the dot dashed line -2.15}
\label{fig:isoplot}
\end{figure}

Because of the lack of reliable Hipparcos parallaxes, the unreliability of log $\textit{g}_{\rm Tphot}$, and concerns about trying to constrain $\textit{T}_{\rm eff}$ and log $\textit{g}$ simultaneously with WIDTH6, we then switched to using the log $\textit{g}_{\rm iso}$ implied by the Yonsei-Yale theoretical isochrones. These isochrones can be seen in Figure \ref{fig:isoplot}. Three different metallicities are plotted; the other available metallicities are [Fe/H] = -2.3 and -2.76. We use $\textit{T}_{\rm phot}$ to obtain log $\textit{g}_{\rm iso}$, and then null the trends of abundance with excitation energy to obtain the excitation energy temperature, $\textit{T}_{\rm \chi}$. In most cases this leads to two scales for temperature: a MS value and a SGB value. Only if the Hipparcos gravity is sufficiently accurate can we be sure of the evolutionary state

An initial comparison between results calculated using the KURUCZ93 models, with overshoot, and those for the new KURUCZ06 models, with no-overshooting, was carried out for the star HD140283. This was done using the calculated Hipparcos gravity for the star, log $\textit{g} = 3.73$. It was found that the KURUCZ06 models gave $\textit{T}_{\chi}$ hotter by 41~ K than the KURUCZ93 models. We decided that for the remainder of the analysis the KURUCZ06 models would be used; see Sect. 3.2 for a discussion on the reasons for this.
\section{Results}
\label{sec:Results}

The physical parameters for the sixteen programme stars and the two standards are given in Table 4.

\begin{table*}[htp]
	\caption{Parameters deduced using the photometric temperatures and the gravities calculated using the Hipparcos parallax. Errors in log $\textit{g}_{\rm Tphot}$ are from the ionisation balance using WIDTH6, whilst those in log $\textit{g}_{\rm Hipp}$ are from the errors in the Hipparcos parallax. The error in log $\textit{g}_{\rm Tphot}$ comes about from the statistical error in making sure the the abundance from different ionisation states is the same, ionisation balance.}
	\label{table:3}
		\centering
		\begin{tabular} {l c c c c c c}
		\hline\hline
		Star name	& No. of \ion{Fe}{i} &$\textit{T}_{\rm phot}$ (K)	&	log $\textit{g}_{\rm Tphot}$	&	[Fe/H]&	$\xi$(Km/s)	&	log $\textit{g}_{\rm Hipp}$	\\
		&lines measured & & & & & \\
		\hline
BD-13$^{\circ}$3442	& 86 &6210 	&	3.77 $\pm$ 0.04	&	$-2.72$	&	1.30	& ...	\\
BD+20$^{\circ}$2030	&	98 &6200 	&	3.985 $\pm$ 0.04	&	$-2.62$	&	1.20	&	...	\\
BD+24$^{\circ}$1676	& 96 &6170 	&	3.68 $\pm$ 0.04	&	$-2.60$	&	1.40	&	...	\\
BD+26$^{\circ}$2621	&	107 &6150 	&	3.88 $\pm$ 0.10	&	$-2.69$	&	1.20	&	4.28 $\pm$ 0.31	\\
BD+26$^{\circ}$3578	&	126 &6150 	&	3.72 $\pm$ 0.08	&	$-2.39$	&	1.35	&	3.94 $\pm$ 0.30	\\
BD+3$^{\circ}$740	&	94 &6240 	&	3.68 $\pm$ 0.12	&	$-2.77$	&	1.50	&	4.18 $\pm$ 0.21	\\
BD+9$^{\circ}$2190	&	146 &6250 	&	3.68 $\pm$ 0.10	&	$-2.84$	&	1.50	&	...	\\
CD-24$^{\circ}$17504	&	69 &6070 	&	3.57 $\pm$ 0.30	&	$-3.35$	&	1.30	& ...	\\
CD-33$^{\circ}$1173	&	126 &6250 	&	3.87 $\pm$ 0.10	&	$-3.04$	&	1.50	&	4.33 $\pm$ 0.27	\\
CD-35$^{\circ}$14849	&	 69 &6060 	&	3.92 $\pm$ 0.07	&	$-2.44$	&	1.20	&	4.44 $\pm$ 0.25	\\
CD-71$^{\circ}$1234	&	127 &6190 	&	3.90 $\pm$ 0.07	&	$-2.54$	&	1.50	&	…	\\
G64-12&	59 &	6220 	&	4.05 $\pm$ 0.08	&	$-3.40$	&	1.40	& ...	\\
G64-37&	71 &	6240 	&	4.29 $\pm$ 0.04	&	$-3.24$	&	1.30	& ...	\\
LP635-14&	163 &	6270 		&3.90 $\pm$ 0.10	&	$-2.49$	&	1.50	& ...	\\
LP815-43&	157 &	6340 		&3.78 $\pm$ 0.12	&	$-2.74$	&	1.40	&	...	\\
HD 84937	& 164	&6160 		&3.66 $\pm$ 0.10	&	$-2.32$	&	1.30	&	3.98 $\pm$ 0.12	\\
\hline\hline
HD 140283	&	124 &5750 	&	3.40 $\pm$ 0.04	&	$-2.53$	&	1.50	&	3.73 $\pm$ 0.12	\\
HD 74000	&	84 &6040 	&	3.77 $\pm$ 0.03	&	$-2.19$	&	1.20	&	4.03 $\pm$ 0.18	\\
\hline
Sensitivity to +100 K in $\textit{T}_{\rm phot}$&  &+0.15& & &  &$\pm$0.04\\
\hline
\end{tabular}
\end{table*}

The effective temperatures for all but the HD stars in Table 4 were found using isochronal gravities, log $\textit{g}_{\rm iso}$, based in turn on photometric temperatures. In order to understand how reliable our $\textit{T}_{\rm \chi}$ determinations are, it is essential that we quantify the various sources of possible error that contribute. As stated above, we defer discussions of non-LTE effects to Paper II. One source of error is the adopted gravity, log $\textit{g}_{\rm Hipp}$, for the HD stars and log $\textit{g}_{\rm iso}$ for the remainder. Choice of isochrones older/younger by 1 Gyr would change log $\textit{g}_{\rm iso}$ by $\sim$0.03 dex for MS and $\sim$0.06~ dex for SGB stars, corresponding to changes in $\textit{T}_{\rm \chi}$ by 12 K and 24 K respectively. The isochronal gravity is also sensitive to the choice of $\textit{T}_{\rm eff}$, for which we use $\textit{T}_{\rm phot}$. An error of +100~ K in $\textit{T}_{\rm phot}$ would produce an error of typically +0.06~ dex in log $\textit{g}_{\rm iso}$ in the SGB gravity, and -0.06~ dex in the MS gravity, though more at the turnoff. This in turn produces an error of typically +24 K in the SGB $\textit{T}_{\rm \chi}$, and -24~ K in the MS $\textit{T}_{\rm \chi}$. Hence, and importantly, the inferred $\textit{T}_{\rm \chi}$ is only weakly dependent on an imperfect value of $\textit{T}_{\rm phot}$. Concerning microturbulence, we see that an error of 0.1~ kms$^{-1}$ leads to, on average, an error of $\approx$60~ K in $\textit{T}_{\rm \chi}$. Typical errors in microturbulence for this work range from $\sim$0.05 - 0.15~km~s$^{-1}$

\begin{table*} [ht]
	\caption{The final physical parameters deduced in this work for the main sequence (MS) scale and the sub giant (SGB) scale. The three HD stars had gravities determined by their Hipparcos parallaxes and as such their evolutionary state can be deduced.}
	\label{table:4}
		\centering
		\begin{tabular} {l c c c c c c c c}
		\hline\hline
		Star name& log $\textit{g}_{\rm iso}$	&	$\textit{T}_{\rm \chi}$ (K)	&	[Fe/H]	&	$\xi$ (Km/s)	&	log $\textit{g}_{\rm iso}$	&	$\textit{T}_{\rm \chi}$ (K)	&	[Fe/H]	&	$\xi$ (Km/s)	\\
		    & (MS) & (MS) & (MS) & (MS) & (SGB) & (SGB) & (SGB) & (SGB) \\ 
		\hline
BD-13$^{\circ}$ 3442	&	4.46 $\pm$ 0.14	&	6321 $\pm$ 87	&	-2.66	&	1.20	&	3.79 $\pm$ 0.14	&	6186 $\pm$ 87	&	-2.74	&	1.30 \\
BD+20$^{\circ}$ 2030	&	4.46 $\pm$ 0.14	&	6208 $\pm$104	&	-2.63	&	1.00	&	3.79 $\pm$ 0.14	&	6099 $\pm$ 104	&	-2.69	&	1.20 \\
BD+24$^{\circ}$ 1676	&	4.47 $\pm$ 0.14	&	6296 $\pm$ 55	&	-2.53	&	1.20	&	3.78 $\pm$ 0.14	&	6220 $\pm$ 55	&	-2.57	&	1.40 \\
BD+26$^{\circ}$ 2621	&	4.49 $\pm$ 0.14	&	6233 $\pm$ 107	&	-2.67	&	1.20	&	...	&	...	&	...	&	... \\
BD+26$^{\circ}$ 3578	&	...	& ...	&	...	&	...	&	3.76 $\pm$ 0.14	&	6148 $\pm$ 81	&	-2.42	&	1.35 \\
BD+3$^{\circ}$ 740	&	4.44 $\pm$ 0.14	&	6344 $\pm$ 113	&	-2.72	&	1.50	&	3.81 $\pm$ 0.14	&	6188 $\pm$ 113	&	-2.82	&	1.50 \\
BD+9$^{\circ}$ 2190	&	4.44 $\pm$ 0.14	&	6486 $\pm$ 129	&	-2.68	&	1.50	&	3.81 $\pm$ 0.14	&	6352 $\pm$ 129	&	-2.75	&	1.50 \\
CD-24$^{\circ}$ 17504	&	4.55 $\pm$ 0.14	&	6102 $\pm$ 232	&	-3.33	&	1.00	&	3.61 $\pm$ 0.14	&	6110 $\pm$ 232	&	-3.31	&	1.30 \\
CD-33$^{\circ}$ 1173	&	4.44 $\pm$ 0.14	&	6386 $\pm$ 55	&	-2.94	&	1.50	&	...	&	...	&	...	&	... \\
CD-35$^{\circ}$ 14849	&	4.52 $\pm$ 0.14	&	6168 $\pm$ 38	&	-2.36	&	1.00	&	...	&	...	&	...	&	... \\
CD-71$^{\circ}$ 1234	&	4.46 $\pm$ 0.14	&	6194 $\pm$ 53	&	-2.52	&	1.20	&	3.78 $\pm$ 0.14	&	6172 $\pm$ 53	&	-2.54	&	1.50 \\
G64-12	&	4.49 $\pm$ 0.14	&	6333 $\pm$ 90	&	-3.32	&	1.30	&	3.67 $\pm$ 0.14	&	6304 $\pm$ 90	&	-3.31	&	1.40 \\
G64-37	&	4.48 $\pm$ 0.14	&	6175 $\pm$ 106	&	-3.29	&	1.20	&	3.69 $\pm$ 0.14	&	6181 $\pm$ 106	&	-3.30	&	1.40 \\
LP635-14	&	4.43 $\pm$ 0.14	&	6319 $\pm$ 114	&	-2.48	&	1.50	&	3.83 $\pm$ 0.14	&	6135 $\pm$ 114	&	-2.61	&	1.50 \\
LP815-43	&	4.40 $\pm$ 0.14	&	6529 $\pm$ 107	&	-2.61	&	1.40	&	3.87 $\pm$ 0.14	&	6400 $\pm$ 107	&	-2.68	&	1.40 \\
HD84937	&	...	&	...	&	... &	...	&	3.98 $\pm$ 0.12	&	6168 $\pm$ 102	&	-2.34	&	1.30 \\
\hline\hline
HD140283	&	...	&	...	&	...	&	...	&	3.73 $\pm$ 0.12	&	5769 $\pm$ 39	&	-2.54	&	1.50 \\
HD74000	&	4.03 $\pm$ 0.18	&	6070 $\pm$ 127	&	-2.20	&	1.20	&	...	&	...	&	...	&	... \\

\hline
\end{tabular}
\end{table*}

\begin{table*} [!ht]
	\caption{The final lithium abundances calculated with the excitation energy temperatures ($\textit{T}_{\rm \chi}$)}
	\label{table:5}
		\centering
		\begin{tabular} {l c c c c c}
		\hline\hline
		Star name	&	EW (Li)	&	$\textit{T}_{\rm eff}$ (K) (MS)	&	$A$(Li) (MS)	&	$\textit{T}_{\rm eff}$ (K) (SGB)	&	$A$(Li) (SGB)	\\
		\hline
BD-13$^{\circ}$ 3442		&	21.0 &	6321 $\pm$ 87	&	2.20 $\pm$ 0.057	&	6186 $\pm$ 87	&	2.10 $\pm$ 0.057	\\
BD+20$^{\circ}$ 2030		&	20.5	&	6208 $\pm$104	&	2.11 $\pm$ 0.068	&	6099 $\pm$104	&	2.03 $\pm$ 0.068	\\
BD+24$^{\circ}$ 1676		&	21.1&	6296 $\pm$ 55	&	2.18 $\pm$ 0.009	&	6220 $\pm$ 55	&	2.13 $\pm$  0.009	\\
BD+26$^{\circ}$ 2621		&	22.5	&	6233 $\pm$ 107	&	2.17 $\pm$ 0.070	&	...	&	...	\\
BD+26$^{\circ}$ 3578		&	24.6	&	...	&	...	&	6148 $\pm$ 81	&	2.15 $\pm$ 0.053	\\
BD+3$^{\circ}$ 740		&	19.5	&	6344 $\pm$ 113	&	2.18 $\pm$ 0.074	&	6188 $\pm$ 113	&	2.07 $\pm$  0.074	\\
BD+9$^{\circ}$ 2190		&	14.6	&	6486 $\pm$ 129	&	2.14 $\pm$ 0.084	&	6352 $\pm$ 129	&	2.05 $\pm$  0.084	\\
CD-24$^{\circ}$ 17504		&	18.1	&	6102 $\pm$ 232	&	1.98 $\pm$ 0.151	&	6110 $\pm$ 232	&	1.98 $\pm$  0.151	\\
CD-33$^{\circ}$ 1173		&	17.2		&	6386 $\pm$ 55	&	2.15 $\pm$ 0.036	&	...	&	...	\\
CD-35$^{\circ}$ 14849		&	28.8	&	6168 $\pm$ 38	&	2.24 $\pm$ 0.025	&	...	&	...	\\
CD-71$^{\circ}$ 1234		&	25.9	&	6194 $\pm$ 53	&	2.21 $\pm$ 0.035	&	6172 $\pm$ 53	&	2.19 $\pm$  0.035	\\
G64-12		&	21.2	&	6333 $\pm$ 90	&	2.21 $\pm$ 0.059	&	6304 $\pm$ 90	&	2.19 $\pm$  0.059	\\
G64-37		&	18.2	&	6175 $\pm$ 106	&	2.03 $\pm$ 0.070	&	6181 $\pm$ 106	&	2.03 $\pm$  0.070	\\
LP635-14		&	20.2	&	6319 $\pm$ 114	&	2.18 $\pm$ 0.074	&	6135 $\pm$ 114	&	2.05 $\pm$  0.074	\\
LP815-43		&	16.1	&	6529 $\pm$ 107	&	2.21 $\pm$ 0.069	&	6400 $\pm$ 107	&	2.13 $\pm$  0.069	\\
HD84937		&	24.9	&	...	&	...	&	6168 $\pm$ 102	&	2.17 $\pm$ 0.066	\\
\hline\hline
HD140283		&	47.9	&	...	&	...	&	5769 $\pm$ 39	&	2.21 $\pm$ 0.025	\\
HD74000		&	22.1	&	6070 $\pm$ 127	&	2.05 $\pm$ 0.083	&	...	&	...	\\
\hline
\end{tabular}
\end{table*}

\begin{figure}
\centering

\psfig{file=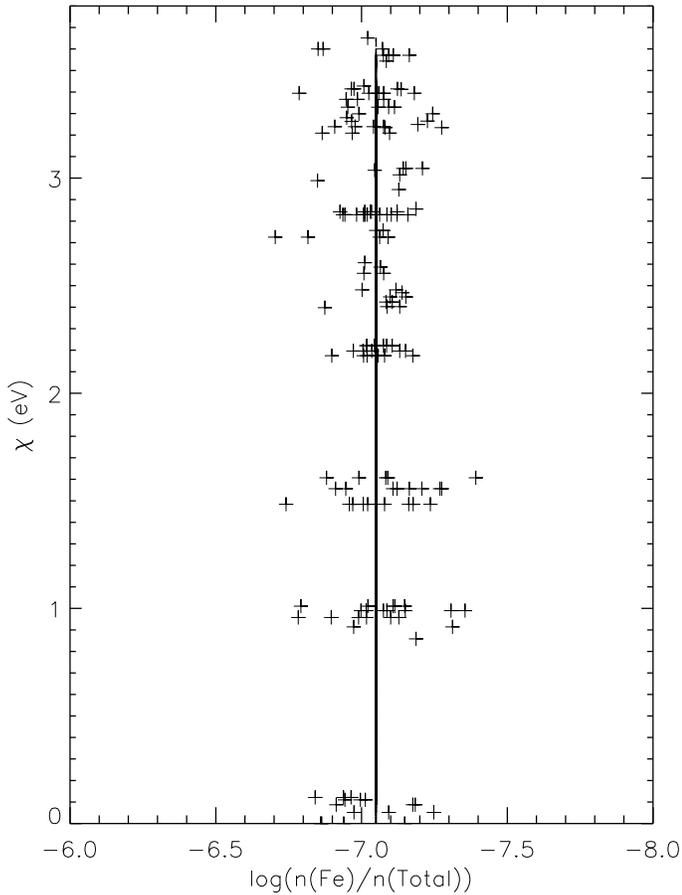,width=9cm}
	\caption{Plot of $\chi$ versus log($\frac{n(Fe)}{n(Total)}$) showing the trend nulling involved in constraining the $\textit{T}_{\rm eff}$.}
\label{fig:chiabund}
\end{figure}

The major contributions to the uncertainty in $\textit{T}_{\rm \chi}$ comes from the procedure of nulling the dependence of [Fe/H] on $\chi$. This involves making a linear least squares fit to the $n$ pairs of $\chi$, [Fe/H] values for a star, using $n$ \ion{Fe}{} lines, as may be seen in Figure \ref{fig:chiabund}. As there is a spread of [Fe/H] values for a given star, the slope can only be determined to a certain statistical accuracy, which turns out to be of order $\sigma \approx$ 0.01 - 0.015 dex per eV, which corresponds to a temperature uncertainty of order $\sim$60 - 80 K. The range in [Fe/H] values encapsulates the impact of random (not $\chi$-correlated) line-to-line errors in equivalent widths, $\textit{gf}$, and damping values. We determine this value for each star, finding a particularly large value for CD-24$^{\circ}$17504 on account of the small number of \ion{Fe}{I} lines in this very metal-poor star. The final 1$\sigma$ uncertainty in $\textit{T}_{\rm \chi}$, given in Table \ref{table:4}, includes the quadrature sum of the (dominant) statistical error arising from nulling the [Fe/H],$\chi$ trend, and contributions for $\Delta$age = 1~ Gyr, $\Delta\xi$ = 0.1 kms$^{-1}$, $\Delta$[Fe/H] = 0.05, and an allowance of $\Delta\textit{T}_{\rm phot}$ = 100 K.

We comment briefly on the larger scale of the uncertainties quoted here, compared to the relatively small values (30 - 40 K) quoted for $\textit{T}_{\rm phot}$ by \cite{Ryanetal99}. The aim of the previous work was to derive temperatures that minimised the impact of  possible systematic errors. The purpose was to investigate the spread of Li abundances by ensuring random errors were kept to a minimum, while at the same time acknowledging that large zero-point errors might persist but be of approximatly the same size in all stars. In the current work, however, the emphasis has been to constrain the zero-point, albeit at the expense of larger random errors for individual stars.

These temperatures have been used in combination with the Li equivalent widths measured by \cite{Ryanetal99}, and a grid of equivalent width versus abundance for different $\textit{T}_{\rm eff}$'s taken from \cite{Ryanetal96a}, to calculate new Li abundances. The grid was constructed by determining different \element[ ][7]{Li} equivalent widths  for a number of different $\textit{T}_{\rm eff}$ and [Fe/H] values. This was done by producing synthetic spectra of the \element[][7]{Li} region with an LTE code originating from \cite{CottrellNorris78}. The log \textit{gf} values and wavelengths of the four components of \element[][7]{Li} were taken from \cite{AndersenGustafssonLambert84} and the model atmospheres were interpolated from the \cite{Bell81} grid. The details of the process can be found in \cite{Norrisetal94}. Table \ref{table:5} presents these values. 
\section{Discussion}
\label{sec:Discussion}

\subsection{Comparison with other temperature scales}
\label{sec:ComparisonWithOtherTemperatureScales}

With the determination of the effective temperatures using the excitation method we can now compare our temperature scale to that of others, in particular to the IRFM of \cite{MelendezRamirez04}, the Balmer line wing method of Asplund et al. (2006), and the principally photometric temperatures of \cite{Ryanetal99}.
\begin{figure} [!h]
	\centering
 \vbox{
		\psfig{file=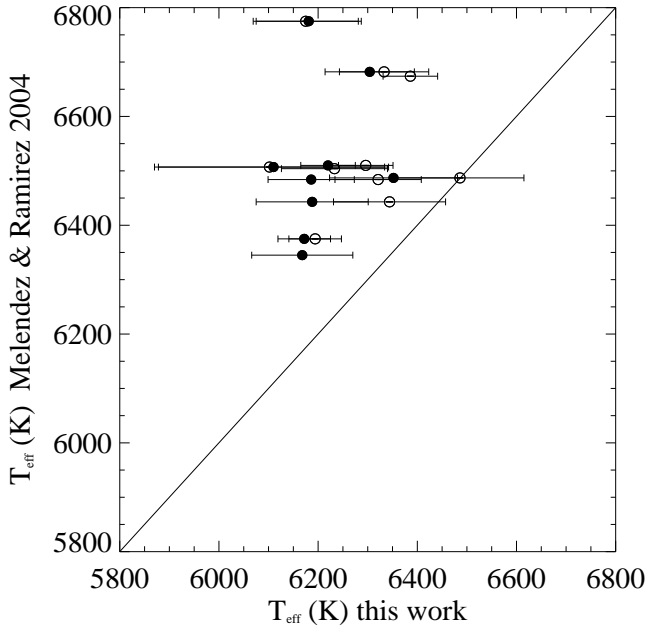,width=8.5cm}
}
	\caption{Comparison of the \cite{MelendezRamirez04} temperatures and the temperatures from this work. Filled circles represent SGB temperatures and open circles represent MS temperatures. Those stars of unknown evolutionary state have two points on the plot.}
	\label{fig:TMR-vs-TAHfnl}
\end{figure}

In Figure \ref{fig:TMR-vs-TAHfnl} the comparison with the IRFM $\textit{T}_{\rm eff}$ of \cite{MelendezRamirez04} is plotted. That scale is clearly hotter than ours, by an average of $\sim$ 250 K. This is not to say that either scale is right, but it does lead us to the conclusion that our temperature scale will not reconcile the Li problem; the Mel\'{e}ndez \& Ram\'{i}rez (2004) IRFM scale was on the borderline of doing so. Our scale will give a lower mean Li abundance than will the IRFM. The difference in temperatures could have several possible explanations. It could be due to problems with our use of LTE in the excitation method, causing it to calculate lower temperatures. An incorrect bolometric flux, $\textit{F}_{\rm bol}$, calibration used in the IRFM could lead to that method having higher temperatures. It has also been noted (Alonso et al. 1996) that the errors in determining the absolute IR flux calibration have different effects on the derived IRFM temperatures, depending on what photometric band is used. The effect of these different errors on the derived temperatures is to move the zero point of the temperature scale. This could be another reason that the \cite{MelendezRamirez04} temperatures are hotter.

\begin{figure} [!ht]
	\centering
		\psfig{file=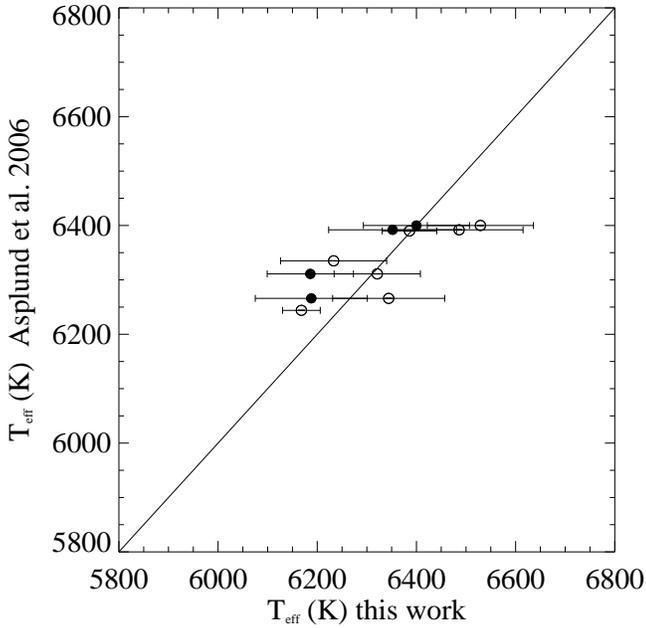,width=8.5cm}
	\caption{Comparison of the \cite{Asplundetal06} temperatures and those from this work. Filled circles represent SGB temperatures and open circles represent MS temperatures. Those stars of unknown evolutionary state have two points on the plot.}
	\label{fig:Tasp-vs-TAHfnl}
\end{figure}

The comparison between the temperatures of \cite{Asplundetal06} and this work is plotted in Figure \ref{fig:Tasp-vs-TAHfnl}. Here we see that the two scales are comparable. As discussed in Sect. 1.2 both of these methods may suffer from LTE effects. A departure from LTE of only a few percent would increase Balmer wing temperatures by of order 100 K (Barklem 2007). It is speculated that the use of LTE in the treatment of the Fe atom may be inaccurate, e.g. overionisation of the element relative to LTE may occur (\cite{Theveninetal99}). If true this would also lead to the excitation temperatures changing, as they are based on an LTE treatment. As yet the value of the temperature change is not determined and the sensitivity of each method to non-LTE  may be very different. This is an aspect that we intend to look into in future work. It is clearly of interest to know whether non-LTE corrections to these scales could increase both to be akin to that of the IRFM scale, and push the mean Li abundances toward that of the WMAP inferred abundances. This agreement between the two scales also gives an indirect comparison with the Balmer wing temperature scale of \cite{Bonifacioetal07}. They found that their temperature scale was essentially the same as \cite{Asplundetal06}, which is in turn similar to ours.
\begin{figure}[!hb]
	\centering
		\psfig{file=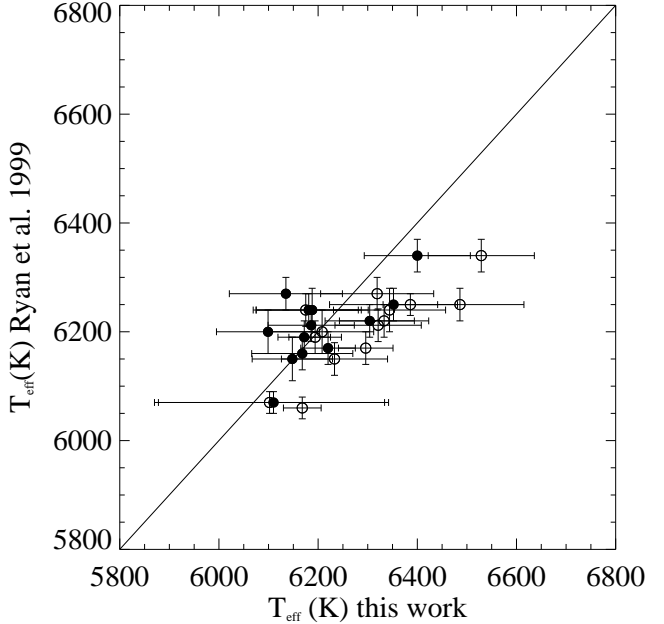,width=8.5cm}
	\caption{Comparison of the \cite{Ryanetal99} temperatures and those of this work. Filled circles represent SGB temperatures and open circles MS temperatures. Those stars of unknown evolutionary state have two points on the plot.}
	\label{fig:Tphot-vs-TAHfnl}
\end{figure}

Finally, Figure \ref{fig:Tphot-vs-TAHfnl} presents the comparison between the photometric temperatures of \cite{Ryanetal99} and $\textit{T}_{\rm \chi}$ of this work. At lower temperatures, $\textit{T}_{\rm \chi} < 6250$ K, the two scales are in excellent agreement. For the MS scale we have seven stars cooler than 6250 K. The difference, $\textit{T}_{\rm \chi} - \textit{T}_{\rm phot}$, ranges from -126 K to 56 K, with a mean of -19 K. For the SGB scales we have eleven stars cooler than 6250 K, with the difference ranging from -99 K to 47 K and a mean of -18 K. However, at higher values, $\textit{T}_{\rm \chi} \geq 6250$ K, we see that the excitation scale becomes hotter. In this case, for eight MS scale stars we have a range of 87 K to 262 K and a mean of 143 K, and for three SGB scale stars a range of 57 K to 117 K with a mean of 73K. This suggests that the $\textit{T}_{\rm phot}$ values are reliable for $\textit{T}_{\rm \chi}$ $\la$ 6250 K, but may be $\sim$70 - 140 K too cool for $\textit{T}_{\rm \chi}$ $\ga$ 6250 K. Photometric temperature errors can be induced through inaccurate dereddening, inaccurate calibration to standards, or a flawed colour-temperature calibration. The last of these is perhaps the most readily encountered; earlier in this work we cited differences of $\sim$100 K between the IRFM scales of Magain (1987) and \cite{Alonsoetal96}. Of course, the difference could also be due to inaccuracies in the model atmospheres used in this work to derive $\textit{T}_{\rm \chi}$.

\subsection{Lithium Abundances}
\label{sec:LithiumAbundances}

Concerning the lithium problem, with our temperature scales we achieve a mean Li abundance of $\textit{A}$(Li)= 2.16 dex for the MS scale and $\textit{A}$(Li)= 2.10 dex for the SGB scale, with a scatter of 0.074 dex and 0.068 dex respectively. The mean Li abundance determined from the five stars whose evolutionary state is defined is $\textit{A}$(Li)= 2.18 dex with a scatter of 0.038 dex. It is pleasing to note the similarity between our mean \element[][7]{Li} abundances and that derived by \cite{Spiteetal96}, $\textit{A}$(Li)= 2.08 ($\pm$0.08) dex, who also use a temperature scale based on iron excitation. \cite{Bonifacioetal07} also found a similar result, $\textit{A}$(Li)= 2.10 ($\pm$0.09) dex, with their Balmer line wing temperature scale.

These values are obviously still too low to reconcile with the WMAP value of $\textit{A}$(Li)$\approx$2.62 dex. Even if we later find that non-LTE corrections give temperature rises of roughly 150 K, we would get mean lithium abundances of $\textit{A}$(Li)$\approx$2.26 dex and $\textit{A}$(Li)$\approx$2.20 dex for the MS and SGB scales respectively. This value is still too low. However, it is not yet known what is the true effect of non-LTE and/or 3D corrections on our temperature scale.

Although the primary purpose of this analysis was to investigate possibly large systematic errors in temperature, we can also investigate whether there is evidence for dependencies of $\textit{A}$(Li) on $\textit{T}_{\rm eff}$ and [Fe/H] within the sample. With the $\textit{A}$(Li) values from Table 5 and the physical parameters calculated for each star we are able to perform a multiple regression fit, linear in $\textit{T}_{\rm eff}$ and [Fe/H]. In these fits we include the stars with known evolutionary state, three on the MS and two on the SGB, combined with the eleven remaining stars of unknown evolutionary state for which we assess two cases: all MS and all SGB. This produces the equation
\begin{center}

	$\textit{A}$(Li)=1.36($\pm$0.84)+0.00019($\pm$0.00013)$\textit{T}_{\rm eff}$
	+0.138($\pm$0.048)[Fe/H] \\
\end{center}
for the MS case and
\begin{center}

	$\textit{A}$(Li)=0.79($\pm$1.22)+0.00025($\pm$0.00019)$\textit{T}_{\rm eff}$
	+0.095($\pm$0.055)[Fe/H] \\

\end{center}
for the SGB case. Here the bracketed values are 1 $\sigma$ errors in the coefficients. The large errors seen in the intercept coefficient comes from the large extrapolation to $\textit{T}$ = 0 and [Fe/H] = 0. We see no statistically significant trend in either  $\textit{T}_{\rm eff}$ or in [Fe/H] for the SGB as each coefficient has errors of similar size to the coefficient itself. Although there is the possibility of a trend existing in [Fe/H] for the MS case (the coefficient is $\sim$3$\sigma$) there is no certainty in this as it is not known whether all the stars belong in the MS evolutionary state. 

The lack of significant trends here, in contrast to the significant [Fe/H] dependence in the study by \cite{Ryanetal96} (and by \cite{Asplundetal06}), comes about because the random uncertainties on the $\textit{T}_{\rm \chi}$ values are much larger than achieved by \cite{Ryanetal99} using photometry. We recall that the primary purpose of the present work was to search for potentially large $\textit{systematic}$ errors in the temperatures, and for that purpose the larger random errors here are tolerable. However, the larger random errors greatly undermine the second use of the dataset.

\begin{figure} [!h]
	\centering
		\psfig{file=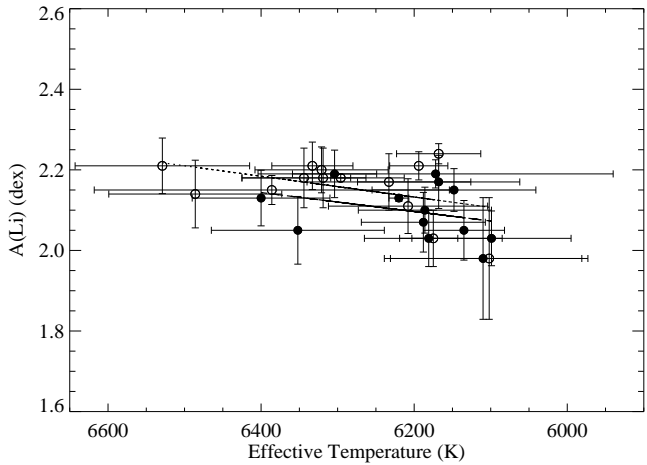,width=8.5cm}
	\caption{Li abundances as a function of temperature. The dotted line is the fit to the MS temperatures (open circles) and the dashed line is the fit to the SGB temperatures (filled circles). Those stars of unknown evolutionary state have two points on the plot.}
	\label{fig:combT-LiABUND}
\end{figure}

\begin{figure} [!h]
	\centering
		\psfig{file=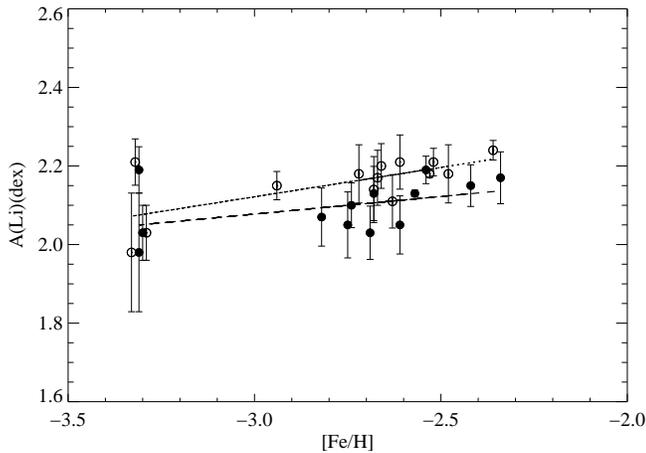,width=8.5cm}
	\caption{Li abundances as a function of metallicity. The dotted line is the fit to the MS metallicities (open circles) and the dashed line is the fit to the SGB metallicities (filled circles). Those stars of unknown evolutionary state have two points on the plot.}
	\label{fig:combmet-LiABUND1}
\end{figure}

We have also performed single parameter regression fits in $\textit{T}_{\rm eff}$ and [Fe/H]. Figure \ref{fig:combT-LiABUND} presents the results of the fits to temperatures. The values of the temperature coefficients imply
\begin{center}

	$\textit{A}$(Li)=0.51($\pm$1.00)+0.00026($\pm$0.00016)$\textit{T}_{\rm eff}$ \\
\end{center}
for the MS case, and
\begin{center}

  $\textit{A}$(Li)=0.71($\pm$1.33)+0.00022($\pm$0.00021)$\textit{T}_{\rm eff}$ \\
\end{center}
for the SGB case. There is clearly no statistically significant trend with temperature in either the MS or the SGB result, again since the temperature coefficient errors are of similar size to the coefficients themselves. For the fit to [Fe/H] we obtained the equation
\begin{center}

		$\textit{A}$(Li)=2.58($\pm$0.14)+0.152($\pm$0.049)[Fe/H]\\
\end{center}
for the MS case, and
\begin{center}

  $\textit{A}$(Li)=2.35($\pm$0.16)+0.089($\pm$0.056)[Fe/H] \\
\end{center}
for the SGB case. Figure \ref{fig:combmet-LiABUND1} shows a plot of these fits. Here we confirm the similarity in values of the metallicity coefficient with \cite{Ryanetal99} and Asplund et al. (2006), but again the large errors for the SGB case do not usefully constrain a metallicity trend. The metallicity coefficient for the MS case is $>3\sigma$, but we caution again that we cannot be sure that all stars are on the MS.

We have also used the fitting form as described by Ryan et al. (2000) to deduce the primordial Li abundance. This fit is of the form:
\begin{equation}
Li/H = a^{'}+b^{'}Fe/Fe_{\sun}
\end{equation}
where $a^{'}$ is the intercept and measures the primordial abundance directly. From this we obtain a value of \element[][7]{Li}/\element[][]{H} = ($1.18 \pm 0.10 )\times10^{-10}$ for the MS case and \element[][7]{Li}/\element[][]{H} = $(1.10 \pm 0.1)\times10^{-10}$ for the SGB case. Comparing this to the value deduced from WMAP via BBN, \element[][7]{Li}/\element[][]{H} = $4.15^{+0.49}_{-0.45}\times10^{-10}$, we once again see we have not reconciled the lithium problem.

We can see from this work that it has not been possible to find a solution to the lithium problem through the use of temperatures derived using the excitation technique. The $\textit{T}_{\rm \chi}$ values are lower than the \cite{MelendezRamirez04} values, and more similar to the H$\alpha$ and photometric temperature estimates. It would seem that one of the biggest outstanding problems is the assumption of LTE in the calculations. As has been stated in previous sections, we have begun our own investigation into the effects of non-LTE on abundance analysis, and therefore the effective temperatures, of this group of stars. 

\section{Conclusions}

The physical parameters of 16 program stars, and two standard stars, have been calculated using WIDTH6. In particular, we have derived excitation energy temperatures by nulling the dependence of abundance on excitation potential for \ion{Fe}{i} lines. We have compared our temperatures with those calculated in other works using different techniques, the IRFM of \cite{MelendezRamirez04}, the Balmer line wing method of \cite{Asplundetal06}, which is similar to that of \cite{Bonifacioetal07}, and the photometric method of \cite{Ryanetal99}.

We have shown that the IRFM scale of \cite{MelendezRamirez04} is hotter than ours by an average of $\sim$250 K. This difference may be the effect of the $\textit{F}_{bol}$ calibration used in IRFM or it could be a shift of the zero-point as an effect of the errors induced by using different photometric bands in the IRFM calculations. On the other hand it may be the effects of assuming LTE on our temperature scale. 
 
It has been found that our methods have produced temperatures comparable with those derived using the Balmer line wing method. Both of these scales are too low to reconcile the lithium problem. It has been noted that the LTE assumptions of the Balmer line wing method may be leading to temperatures that are on order of 100 K too low. The method used in this work may also suffer from problems with LTE, but the full effect of non-LTE on the excitation method used here is unknown. This is something we will research in future work.

Finally we also see comparable results with the photometric temperature scale, although at higher temperatures, $\textit{T}_{\rm \chi} \geq 6250$ K, there is a shift towards our scale becoming hotter. This suggests that the photometric temperatures are reliable up until $\sim$6250 K. The most readily detected error in photometric calculations comes from the colour-temperature calibration and differences in values derived from different calibrations can be as much as $\sim$100 K.

With our derived temperatures we have calculated the Li abundances of the program stars. Due to large uncertainties, we do not see statistically compelling trends in $\textit{A}$(Li) with either $\textit{T}_{\rm eff}$ or [Fe/H]. Due to the uncertainty in evolutionary state for the majority of the stars we have two mean lithium abundances, $\textit{A}$(Li) = 2.16 dex for the main sequence case and $\textit{A}$(Li) = 2.10 dex for the SGB case. Five of the stars do have defined evolutionary states. The mean Li abundance from these five stars is $\textit{A}$(Li)= 2.12 dex. These values are lower than the WMAP determined value of $\textit{A}$(Li)$\approx$2.62 dex. For the observationally deduced value to be comparable with the WMAP value our temperatures would need to increase by approximately 700 K.

\begin{acknowledgements}

We thank J. D. Tanner for assisting with collecting the initial data at the AAT and the referee, P. Bonifacio, for his helpful comments which have improved the presentation of this work.
The work of KAO was supported in part by DOE Grant No.\ DE-FG02-94ER-40823. AH, SGR, AEGP gratefully acknowledge support from the Royal Society under International Joint Project 2006/23 involving colleagues at Uppsala University.

\end{acknowledgements}

\end{document}